\documentclass[twocolumn]{aastex631} % THIS IS FOR ARXIV SUBMISSION (no line numbers)

\newcommand{\A}{\text{\normalfont\AA}}

\def \spherex{SPHEREx}

\def \ebmv{E(B$-$V)}

\def \logm{\log(\rm M/{\rm M_\odot})}

\def \h2{{\rm H_{2}}}

\def \dn4000{D_{{\rm n}}(4000) }

\begin{document}

\title{The Potential of the SPHEREx Mission for Characterizing Polycyclic Aromatic Hydrocarbon 3.3\,$\mu$m Emission in Nearby Galaxies}

\author[0009-0003-1284-3507]{Edward Zhang} % need to create an ORCID
\affiliation{California Institute of Technology, 1200 E. California Boulevard, Pasadena, CA 91125, USA}
\correspondingauthor{Edward Zhang}
\email{ezhang3@caltech.edu}

\author[0000-0002-9382-9832]{Andreas L. Faisst}
\affiliation{Caltech/IPAC, 1200 E. California Boulevard, Pasadena, CA 91125, USA}

%% Core people (alphabetical)

\author[0000-0002-4650-8518]{Brendan P. Crill}
\affiliation{Jet Propulsion Laboratory, California Institute of Technology, 4800 Oak Grove Drive, Pasadena, CA 91109, USA}

\author[0000-0003-4268-0393]{Hanae Inami}
\affiliation{Hiroshima Astrophysical Science Center, Hiroshima University, 1-3-1 Kagamiyama, Higashi-Hiroshima, Hiroshima 739-8526, Japan}

\author[0000-0001-8490-6632]{Thomas Lai}
\affiliation{Caltech/IPAC, 1200 E. California Boulevard, Pasadena, CA 91125, USA}

\author[0000-0001-9490-3582]{Youichi Ohyama}
\affiliation{Academia Sinica Institute of Astronomy and Astrophysics (ASIAA), No. 1, Section 4, Roosevelt Road, Taipei 10617, Taiwan}

\author[0000-0001-9937-8270]{Jeonghyun Pyo}
\affiliation{Korea Astronomy and Space Science Institute, 776 Daedeokdae-ro, Yuseong-gu, Daejeon 34055, Korea}

%% other contributors (alphabetical)

\author[0000-0001-9674-1564]{Rachel Akeson}
\affiliation{Caltech/IPAC, 1200 E. California Boulevard, Pasadena, CA 91125, USA}

\author[0000-0002-3993-0745]{Matthew L.\ N.\ Ashby}
\affiliation{Center for Astrophysics $|$ Harvard \& Smithsonian, 60 Garden Street, Cambridge, MA 01720, USA}

\author[0000-0002-5710-5212]{James J. Bock}
\affiliation{California Institute of Technology, 1200 E. California Boulevard, Pasadena, CA 91125, USA}
\affiliation{Jet Propulsion Laboratory, California Institute of Technology, 4800 Oak Grove Drive, Pasadena, CA 91109, USA}

\author[0000-0002-5437-0504]{Yun-Ting Cheng}
\affiliation{California Institute of Technology, 1200 E. California Boulevard, Pasadena, CA 91125, USA}
\affiliation{Jet Propulsion Laboratory, California Institute of Technology, 4800 Oak Grove Drive, Pasadena, CA 91109, USA}

\author[0000-0001-6320-261X]{Yi-Kuan Chiang}
\affiliation{Academia Sinica Institute of Astronomy and Astrophysics (ASIAA), No. 1, Section 4, Roosevelt Road, Taipei 10617, Taiwan}

\author[0000-0002-3892-0190]{Asantha Cooray}
\affiliation{Department of Physics \& Astronomy, University of
California, Irvine, CA 92697}

\author[0000-0001-7432-2932]{Olivier Dor{\'e}}
\affiliation{Caltech/IPAC, 1200 E. California Boulevard, Pasadena, CA 91125, USA}
\affiliation{Jet Propulsion Laboratory, California Institute of Technology, 4800 Oak Grove Drive, Pasadena, CA 91109, USA}

\author[0000-0002-9330-8738]{Richard M. Feder}
\affiliation{Berkeley Center for Cosmological Physics, University of California, Berkeley, CA 94720,
USA}
\affiliation{Lawrence Berkeley National Laboratory, Berkeley, CA 94720, USA}

\author[0000-0003-1647-3286]{Yongjung Kim}
\affiliation{School of Liberal Studies, Sejong University, 209 Neungdong-ro, Gwangjin-Gu, Seoul 05006, Republic of Korea}
\affiliation{Korea Astronomy and Space Science Institute, 776 Daedeokdae-ro, Yuseong-gu, Daejeon 34055, Korea}

\author[0000-0003-1954-5046]{Bomee Lee}
\affiliation{Korea Astronomy and Space Science Institute, 776 Daedeokdae-ro, Yuseong-gu, Daejeon 34055, Korea}

\author[0000-0001-5382-6138]{Daniel Masters}
\affiliation{Caltech/IPAC, 1200 E. California Boulevard, Pasadena, CA 91125, USA}

\author[0000-0002-6025-0680]{Gary Melnick}
\affiliation{Center for Astrophysics $|$ Harvard \& Smithsonian, 60 Garden Street, Cambridge, MA 01720, USA}

\author[0000-0002-5158-243X]{Roberta Paladini}
\affiliation{Caltech/IPAC, 1200 E. California Boulevard, Pasadena, CA 91125, USA}

\author[0000-0003-4990-189X]{Michael W. Werner}
\affiliation{Jet Propulsion Laboratory, California Institute of Technology, 4800 Oak Grove Drive, Pasadena, CA 91109, USA}

%\author{SPHEREx team}
%\affiliation{}



\begin{abstract}
Together with gas, stars, and supermassive black holes, dust is crucial in stellar and galaxy evolution. Hence, understanding galaxies' dust properties across cosmic time is critical to studying their evolution. 
In addition to photometric constraints on the absorption of blue light and its reemission at infrared wavelengths, dust grain properties can be explored spectroscopically via polycyclic aromatic hydrocarbon (PAH) emission bands in the mid-IR.
The new SPHEREx space telescope conducts an all-sky spectrophotometric survey of stars and galaxies at wavelengths of 0.75--5$\,\mu$m, making it ideal for studying the widespread presence of the 3.3$\,\mu$m PAH emission across galaxy populations out to $z\sim0.4$.
In this paper, we simulated galaxy spectra to investigate SPHEREx's capability to study PAH emission in such galaxies. We find that for the all-sky survey the PAH 3.3$\,\mu$m emission band flux can be measured to 30\% accuracy at $\logm>9.5$ and star formation rate (SFR) $> 1\,{\rm M_\odot\,yr^{-1}}$ at $z=0.1$, $\logm > 10.5$ and ${\rm SFR} > 10\,{\rm M_\odot\,yr^{-1}}$ at $z=0.2-0.3$, and $\logm>11$ and ${\rm SFR} > 100\,{\rm M_\odot\,yr^{-1}}$ at $z=0.4$. For deep SPHEREx fields, a factor of $\sim 10$ deeper sensitivity limits can be reached.
Overall, SPHEREx will enable the measurement of the 3.3$\,\mu$m PAH band emission in several hundred thousand galaxies across the sky, providing a population study of the smallest dust grains (``nano grains'') and radiation properties in massive galaxies in the nearby Universe.

\end{abstract}

%% https://astrothesaurus.org/concept-select/


\section{Introduction} \label{sec:intro}

Dust is inherently connected to galaxy evolution, providing crucial information on the interstellar medium (ISM) of galaxies, their stellar populations, feedback mechanisms, and even the coevolution between supermassive black holes (SMBHs) and the galaxies in which they typically reside. At the same time, dust is important to be understood for more robust measurements of global and resolved properties of the galaxies, such as total star formation rates (SFR) or metallicities.
The production and destruction of dust are tied to supernovae, the abundance of asymptotic giant branch stars, and metallicity, which influences the formation of dust in the ISM \citep[see][]{valiante09,dwek11,asano13,feldmann15,li20}. Furthermore, winds from young stars or the strong ultraviolet (UV) radiation from an accreting SMBH, also known as the active galactic nucleus (AGN), can contribute to the destruction or dissociation of dust grains and even their ejection out of the ISM \citep[e.g.,][]{lai23}.

%Many galactic mechanisms are involved in setting the abundance of dust in galaxies \citep{REF}. These include the production and destruction of dust via supernovae, the dust production related to asymptotic giant branch (AGB) stars, or dust growth in the interstellar medium (IGM). Furthermore, winds from young stars or the strong radiation from an accreting central supermassive black hole, also known as Active Galactic Nucleus (AGN), can contribute to the destruction and expulsion of dust grains \citep{REF}.
%The dust content and properties in galaxies is therefore strongly correlated with their formation of stars (early stars preferentially form in dusty clouds), the properties of their ISM (such as metal enrichment), and the prevalence of black holes and AGN in galaxies. As such, the study of dust is an important avenue to understand the formation and evolution of galaxies as a whole.

%Many galactic mechanisms are involved in the formation and destruction of dust, such as supernova explosions which can both expel and destroy dust, winds from stars, and strong radiation from the central supermassive black hole known as the Active Galactic Nuclei (AGN).

Due to the wavelength-dependent absorption and scattering of blue light by interstellar dust, imaging and spectroscopic studies at infrared (IR) wavelengths have proven to be instrumental to understand dust. To survey dust in a large sample of galaxies of different properties at high redshifts, photometric measurements (such as the observation of the UV continuum slope or total IR emission) have shown to be extremely efficient \citep[e.g.,][]{bouwens09,scoville15}. However, spectroscopy is essential to understand dust (and specifically the dust grain properties) in more detail to inform models of dust formation and destruction.

Some of the most important dust-related spectroscopic features are the emission bands of polycyclic aromatic hydrocarbon (PAH) molecules at $3-30\,{\rm \mu m}$ \citep{leger84,allamandola85,smith04,werner04}, which can contribute up to $20\%$ to the total IR luminosity \citep{smith07} and $15\%$ to the cosmic carbon budget \citep{draine07}.
The abundances of these strong emission-line bands in galaxy spectra are sensitive to the dust grain size distribution. For example, the PAH $3.3\,{\rm \mu m}$ emission traces neutral small-size grains, while the $7.7$ and $11.3\,{\rm \mu m}$ PAH emission is more sensitive to larger ionized and neutral dust grains, respectively \citep{allamandola99,croiset16,maragkoudakis18,maragkoudakis20}. 
The grain size distribution is affected and altered by various dust formation and destruction mechanisms \citep[see review in][]{li20}. The nano grains traced by the PAH $3.3\,{\rm \mu m}$ feature in particular are highly sensitive to the amplitude and hardness of the ambient radiation field, which makes this feature a diagnostic tool for physical conditions in galaxies' ISM \citep{desert90,schutte93,draine03,draine07}.

Although UV and IR continuum photometry as a means of estimating dust attenuation can both be efficiently conducted in terms of telescope time, spectroscopy of PAHs is challenging, especially at high redshifts.
The Short Wavelength Spectrometer on board the Infrared Space Observatory \citep[ISO;][]{kessler96} offered for the first time the sensitivity and wavelength range ($2.4-45\,{\rm \mu m}$) to study PAH emission spectroscopically in extragalactic sources \citep{helou00,spoon00,peeters04}.
The more sensitive Infrared Spectrograph \citep[$5.2-38\,{\rm \mu m}$;][]{houck04} on board the {\it Spitzer} Space Telescope \citep{werner04b} allowed the exploration of even fainter and more distant PAH emission and its dependence on the local environment in galaxies, as well as their metal abundances \citep[e.g.,][]{engelbracht05,smith07,teplitz07,engelbracht08,galliano08,gordon08,sajina09,hunt10,sandstrom12,pope13}.
The combination with observations of the Infrared Camera \citep[]{onaka07} at $2.5-5\,{\rm \mu m}$, $5-13\,{\rm \mu m}$, or $12-26\,{\rm \mu m}$ on board the AKARI Space Telescope \citep{murakami07} enables the study of the major PAH emission bands, including the $3.3\,{\rm \mu m}$ feature and the strong mid-IR bands \citep[e.g.,][]{imanishi10,ohyama18}.
With that, the PAH emission of approximately $250$ star-forming and luminous IR  galaxies (LIRGs) has been studied by combining Spitzer and AKARI spectroscopy \citep[e.g.,][]{inami18,lai20}. For a detailed review on the achievements of Spitzer and AKARI regarding PAH studies, see \citet{li20}.

Now with the {\it James Webb} Space Telescope \citep[JWST;][]{rigby23} up and running, the PAH emission can be spectroscopically detected out to $z=4$ with the MIRI instrument and even spatially resolved with the help of strong gravitational lensing \citep{lai22,spilker23,young23}.
Furthermore, a recent study by \citet{lyu25} measures the PAH $3.3\,{\rm\mu m}$ and aliphatic emission out to $z=0.5$ for a large sample of $187$ galaxies at lower stellar masses.
While this allows a detailed study of dust in individual galaxies, the lack of sky coverage due to JWST's small field of view limits the statistical power. Narrowband or medium-band imaging surveys with JWST may reveal PAH emission at lower spectral resolution but deeper sensitivity for larger samples \citep{rieke24,shivaei24}. However, in these cases it is difficult to spectrally isolate the PAH emission from the continuum, and its measurement is inherently uncertain owing to varying contributions of silicate absorption.
These studies so far have shown the tight correlation between PAH emission, star forming properties, metal abundances, and AGN. However, while these observations are extremely valuable to understand the details of PAH emission, they lack the number statistics for a population-averaged assessment.

% Talk about SPHEREx.
The launch of the NASA MIDEX mission {\it Spectro-Photometer for the History of the Universe, Epoch of Reionization, and Ices Explorer} \citep[\spherex;][]{crill20}\footnote{\url{https://spherex.caltech.edu}} on 2025 March 11 opened up a new avenue to study the PAH $3.3\,{\rm \mu m}$ emission out to $z\sim0.4$. \spherex~will carry out an {\it all-sky} survey, resulting in redshift measurements of $\sim400$ million extragalactic sources at wavelengths of $0.75-5\,{\rm \mu m}$. It will be equivalent to a low-resolution spectral survey at $R\sim40-130$ in $\sim102$ spectral channels.
This wavelength range is split into six detectors with a per-channel sensitivity of $\sim19.5\,{\rm mag}$ (detectors $1-5$) and $\sim18.5$ mag (detectors $5-6$ starting at $3.8\,{\rm \mu m}$).
%The average depth per channel is between $19$ and $22\,{\rm mag}$ (AB) at $5\sigma$ for a point source, depending on how many times a source is visited during the two-year nominal mission time. For example,
Sources close to the north ecliptic pole (NEP) and south ecliptic pole (SEP) are visited $\sim 50-100\times$ more frequently during the $2$ yr nominal mission, resulting in a $\sim 7-10$-fold increase in depth\footnote{Note that only the central $r\sim1.5\,{\rm deg}$ parts of the NEP and SEP receive $100\times$ the visits of the wide-field survey.}. We refer to these fields as the ``deep-field'' survey in the following, while the other survey area is referred to as the ``wide-field'' survey. For more details on the survey strategy of \spherex~, we refer to \citet{bryan25}.
Similar to previous programs with ISO, Spitzer, and AKARI, \spherex~is measuring the PAH emission at higher spectral resolution, therefore mitigating the challenges of JWST/MIRI narrow- and medium-band observations. However, the all-sky coverage of \spherex~allows such measurements for significantly larger samples.

% Now summarize what we are planning to do in this paper.
In this paper, we study the capabilities of \spherex~in observing the PAH $3.3\,{\rm \mu m}$ emission feature, an important indicator of small dust grains and the radiation environment, in galaxies out to $z=0.4$. Given the $5\,{\rm \mu m}$ cutoff, $z=0.4$ is a reasonable upper limit in redshift at which this PAH spectroscopic feature can be observed by \spherex's wavelength coverage. We calibrate a physical model and use the \spherex~observing plan and sensitivity limits to investigate the parameter space in stellar mass, SFR, dust abundance, and redshift in which \spherex~is able to robustly measure this spectroscopic dust grain feature.

This paper is organized as follows: In Section~\ref{sec:datamodel}, we outline the data and calibration of our model to predict the PAH $3.3\,{\rm \mu m}$ as a function of basic galaxy properties (stellar mass, SFR, and dust attenuation). In Section~\ref{sec:sims}, we simulate realistic \spherex~spectra from templates created by our calibrated model. The results are discussed in Section~\ref{sec:results} and we conclude in Section~\ref{sec:conclusions}. 
Throughout this work, we assume a $\Lambda$CDM cosmology with $H_0 = 70\,{\rm km\,s^{-1}\,Mpc^{-1}}$, $\Omega_\Lambda = 0.7$, and $\Omega_{\rm m} = 0.3$ and magnitudes are given in the AB system \citep{oke74}. We use a \citet{chabrier03} initial mass function (IMF) for stellar masses and SFRs.

\section{Data \& Model} \label{sec:datamodel}

The goal of this work is to carry out realistic spectral simulations to characterize the parameter space in which \spherex~will significantly detect the PAH $3.3\,{\rm \mu m}$ feature in galaxies.  To this end, we create realistic galaxy spectral templates and calibrate them to observations. We outline the process below.

\subsection{Models}

We use the {\it Code Investigating Galaxy Emission} \citep[\textsc{CIGALE};][]{burgarella05,noll09,boquien19} to create the spectral models.
We assume that the galaxies have been forming stars for the past $8\,{\rm Gyr}$ at a constant star formation history (SFH). This smooth star formation assumes that the galaxies are following the star-forming main sequence \citep{daddi07,schreiber15}; however, we will also investigate specific SFHs that resemble a starburst with briefly enhanced star formation activity.
The redshift is restricted to $z<0.4$, set by the observability of the PAH $3.3\,{\rm \mu m}$ band by \spherex. The SFR is treated as a free parameter (mimicking different SFRs at different redshifts), and the stellar mass is simply the product of the age and the constant SFR. Note that we assume that galaxies are star forming, as the dust properties of quiescent galaxies\footnote{Defined as galaxies with no recent star formation.} have not yet been conclusively constrained; however, there are hints of them having significantly lower dust abundances \citep{whitaker21,donevski23,michalowski24,lee24}. Quiescent galaxies are therefore not part of our model and beyond the scope of this paper --- but they could be investigated with \spherex~in a stacking analysis (see Section~\ref{sec:conclusions}).
We investigate the effect of starburst galaxies and highly dust-obscured galaxies on the PAH emission with respect to our constant SFH model further in  Appendix~\ref{app:differentmodel}.
%Due to the assumption that the galaxies are star forming on the main sequence, the details of the star formation history (SFH), such as a bursty or delayed SFHs, do not change the results of this paper significantly.

The underlying spectral templates are based on \citet{bruzual03} models assuming a \citet{chabrier03} IMF. We assume solar gas phase and stellar metallicity, which is reasonable for massive galaxies at the redshifts probed here \citep[see  measurements for AKARI galaxies;][]{oi17}, and use an ionization parameter of $\log(U) = -2$ and an electron density of $100\,{\rm cm^{-3}}$ \citep[consistent with measurements; e.g.,][]{isobe23}.
Note that changing the metallicity by $90\%$ to $0.1$ solar would decrease the PAH $3.3\,\mu m$ flux by less than $25\%$. We note that this is less than the factors of a few observed in \citet{engelbracht08} but consistent with the change in {\it total} PAH emission derived in \citet{whitcomb24}.\footnote{ However, we advise caution in interpreting this comparison because the here-quoted $25\%$ is for an ``unphysical'' change in {\it only} metallicity without changing, for example, the total dust mass or other physical parameters of the galaxy, which will impact the PAH emission.}
Changing the ionization parameter and electron density would decrease the flux by $\sim 1\%$. No contribution of AGN is assumed in the following, which could significantly weaken and even remove any trace of the PAH $3.3\,{\rm \mu m}$ band emission \citep[e.g.,][]{lai23}. The effect of AGN is further investigated in Appendix~\ref{app:differentmodel}.
We assume a $100\%$ escape fraction for emission lines (which affects primarily the optical emission and not the PAH emission) and a differential dust attenuation factor between stellar continuum and nebular lines set to $0.44$ \citep{calzetti00}. The strength of the PAH $3.3\,{\rm \mu m}$ band is determined by the level of star formation, the dust abundance (here implemented by the reddening in the \ebmv~value), and $q_{\rm PAH}$, which is the mass fraction (to total dust mass) of the PAHs. 
For the attenuation model, we apply the \texttt{dustatt\_modified\_starburst} module, which is based on an empirical dust attenuation law based on \citet{calzetti00} and extended with the \citet{leitherer02} curve between the Lyman break and $1500\,{\rm \AA}$. The $2175\,{\rm \AA}$ absorption feature (``UV bump'') can be highly correlated with PAH band emission \citep{shivaei22}. The bump is parameterized with a Lorentzian-like Drude profile similar to \citet{noll09}, and we assume a bump amplitude of $3$ and width of $35\,{\rm \AA}$, consistent with the Milky Way attenuation law. However, changing these parameters in reasonable limits results in PAH $3.3\,{\rm \mu m}$ flux changes of less than $2\%$ for a given \ebmv, which is well within the requirements for this experiment.
For dust emission we use the \texttt{dl2014} module based on \citet{draine14}. This is a modified version of the \citet{draine07} model, including a wider range of radiation field intensities and PAH mass fractions ($q_{\rm pah}$), a variable mid-IR power-law index ($\alpha_{\rm mir}$), a change in the treatment of graphite, and renormalized dust masses. We use standard parameters such as $\alpha_{\rm mir}=2$ and let $q_{\rm pah}$ be a free parameter to reproduce the observations (see Section~\ref{sec:calibration}).
We note that in the following we assume that all foreground dust attenuation in our Milky Way has been subtracted robustly in the observations; hence, it is not included in our model.

%%% FIGURE 1 %%%%
\begin{figure}[t!]
%\vspace{5cm}
\centering
\includegraphics[angle=0,width=0.99\columnwidth]{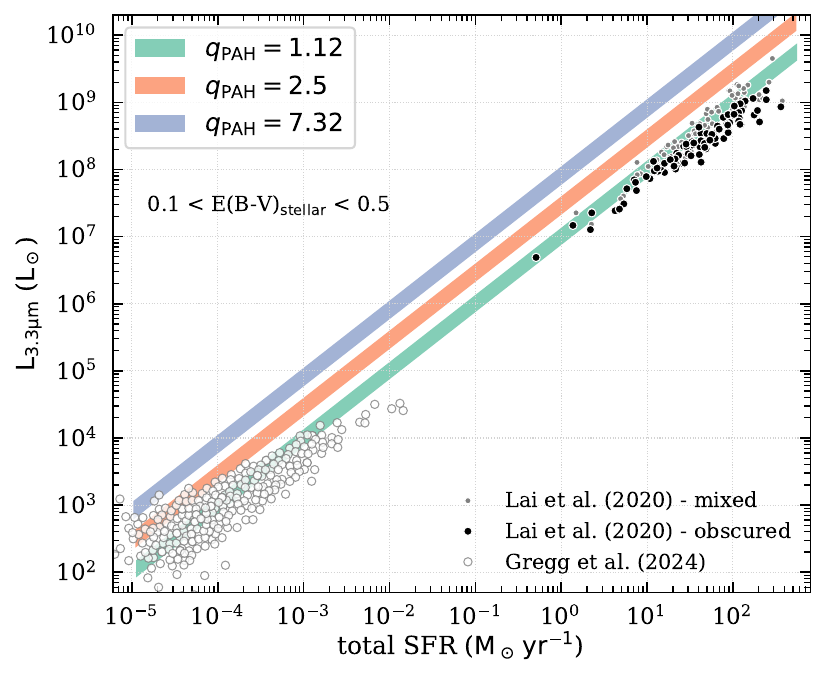}\\
\includegraphics[angle=0,width=0.99\columnwidth]{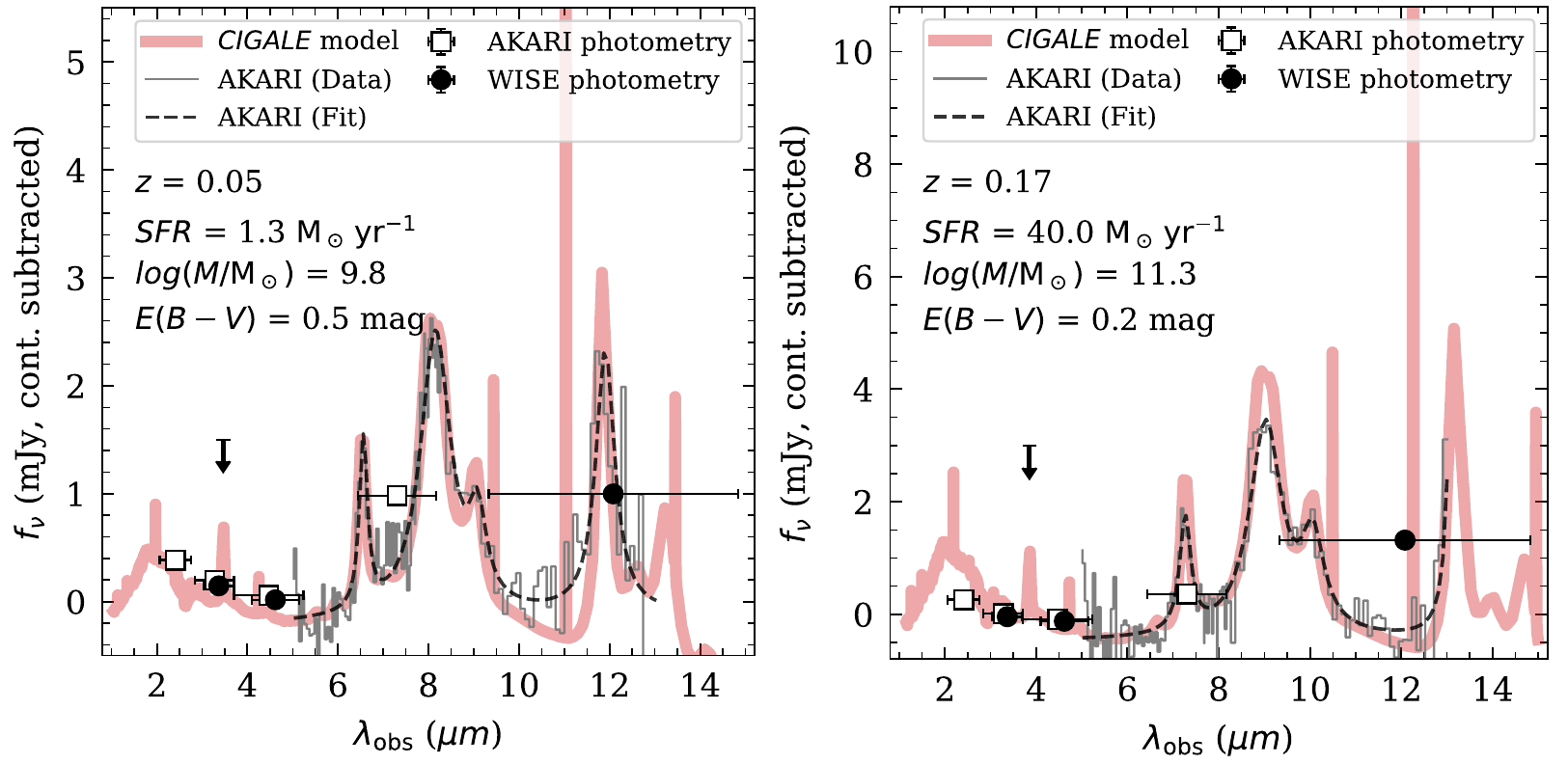}
\caption{{\it Top:} calibration of our \textsc{CIGALE} models to observations from \citeauthor{gregg24} (\citeyear{gregg24}; individual star forming regions in local galaxies) and \citeauthor{lai20} (\citeyear{lai20}; low-redshift star forming galaxies). The models are shown for different $q_{\rm PAH}$ (acting as free parameter, indicated by colors) and a range of \ebmv~from $0.1$ to $0.5\,{\rm mag}$ (indicated by the width of the bands). We find that a value of $q_{\rm PAH} = 1.12\%$ fits well the observations across more than 7 orders of magnitude in total SFRs.
{\it Bottom:} comparison of our \textsc{CIGALE} models (red) to observed AKARI SPICY spectra (data and best fit) from \citet{ohyama18}. Stellar mass is fixed to the one measured from photometric SED fitting, and a $q_{\rm PAH} = 1.12\%$ is assumed. Note that the AKARI spectra cover a wavelength range redder than \spherex~(specifically the 6.2, 7.7, 8.6, 11.3, and $12.7\,{\rm \mu m}$ PAH bands) and therefore miss the $3.3\,{\rm \mu m}$ band (indicated by the arrow). We also show photometric data from WISE (filled squares) and AKARI (open circles) for comparison. 
\label{fig:cigalecalibration}}
\end{figure}

\subsection{Calibration of the Model} \label{sec:calibration}

Having set up the model, we now calibrate it to observations.
The top panel of Figure~\ref{fig:cigalecalibration} compares our model-derived PAH $3.3\,{\rm \mu m}$ luminosity and total SFRs for different PAH mass fractions (color-coded) to observations from \citeauthor{gregg24} (\citeyear{gregg24}; individual star forming regions in local galaxies) and \citeauthor{lai20} (\citeyear{lai20}; low-redshift star forming galaxies). The latter is derived assuming two different attenuation geometries (mixed and obscured; see \citealt{lai20} for more details), however, the difference between these models is minimal and does not affect our calibration. The thickness of the colored areas corresponds to varying stellar continuum dust attenuation, given by \ebmv~values ranging from $0.1$ to $0.5\,{\rm mag}$. In the following we use $q_{\rm PAH}$ as a free parameter to calibrate our model. We find that a value of $1.12\%$ fits best the observations over a range of more than 7 orders of magnitude in total SFR.
We note that this model also successfully fits ultraluminous IR galaxies (ULIRGs), which are included in the samples shown in Figure~\ref{fig:cigalecalibration}. As \spherex~will, due to its sensitivity limits, mostly detect highly star forming galaxies (Section~\ref{sec:results}), we are confident about our assumptions.
The scatter in the PAH vs. SFR relation might be due to potentially different SFHs or lower/higher dust attenuation \ebmv~values. These differences are, however, within a factor of $\sim 2$ and we therefore, without loss of generality, use a value of $q_{\rm PAH} = 1.12\%$ for the remainder of this paper (but we will discuss the impact of other $q_{\rm PAH}$ values).

In addition, we compare the \textsc{CIGALE} model directly to the observed spectra by the {\it slitless spectroscopic survey of galaxies} \citep[SPICY;][]{ohyama18} observed with AKARI's MIR-S camera covering $5-13\,{\rm \mu m}$. The SPICY sample contains $54$ galaxies in the NEP observed with ISO showing a flux excess at $15\,{\rm \mu m}$. Because the spectral coverage between SPICY and \spherex~does not overlap, we cannot directly compare the PAH $3.3\,{\rm \mu m}$ band, but only the spectral features at 6.2, 7.7, 8.6, 11.3, and $12.7\,{\rm \mu m}$. We match the SPICY galaxies to the Subaru Hyper Suprime-Cam (HSC) photometry catalog covering the AKARI NEP field from \citet{kim21}. This catalog contains spectral energy distribution (SED) fits to the HSC photometry and provides stellar mass and SFR rate measurements, as well as photometry from UV to far-IR.
The bottom panel of Figure~\ref{fig:cigalecalibration} shows the spectra together with the Wide-field Infrared Survey Explorer (WISE) and AKARI photometry \citep{kim21} of two SPICY galaxies at $z=0.05$ and $z=0.17$ with overlaid \textsc{CIGALE} models. (Note the different stellar mass and SFR of the two example galaxies.) We fix the stellar mass of our models to the value obtained from the HSC photometry, and we find that the SFR of the model is slightly higher than the one derived from SED fitting in the \citet{kim21} catalog. This could indicate some dust-obscured star formation that is not accounted for in the SED fit or differences in the SFH (note that we are assuming a constant SFH, but see also Appendix~\ref{app:differentmodel}). In addition, the models assume $q_{\rm PAH} = 1.12\%$ and the \ebmv~value is consistent with the value obtained from SED fitting. This test shows that the \textsc{CIGALE} models are reasonably predicting the redder PAH emissions, and we therefore conclude that also the PAH $3.3\,{\rm \mu m}$ emission is predicted well, consistent with the test above.

%We first compare the total PAH $3.3\,{\rm \mu m}$ luminosity to the model and observed SFRs and then the overall PAH shapes to {\it AKARI} and {\it Spitzer} observed spectra.

% explain CIGALE
% explain calibration to SFR vs. PAH 3.3um relation
% compare with AKARI spectrum (of 7um)
% assumption on SFR (mass?)

% Describe catalogs.

% explain we used
% explain model
% explain calibration

%This is the data section. 
%
%\begin{itemize}
%    \item Explain the data that we are using from {\it Spitzer}
%    \item Explain the data that we are using from AKARI
%    \item Explain the selection of the galaxies (based on the \spherex~wavelength and sensitivity)
%    \item Explain the {\it Hyper Suprime-Cam} (HSC) catalog and matching of all the above.
%    \item also add how you measured the stellar masses.
%    \item maybe add a figure showing the final sample (e.g., some SEDs? Magnitude and redshift distributions?)
%    \item we should also have a plot here on what sources are observed how many times with \spherex. Probably a ``aitoff'' projection with the sources and underlying the number of visits by \spherex.   
%\end{itemize}

%%% FIGURE 2 %%%%
\begin{figure}[t!]
%\vspace{5cm}
\centering
\includegraphics[trim={2.5mm 0 0 0},clip,angle=0,width=1.0\columnwidth]{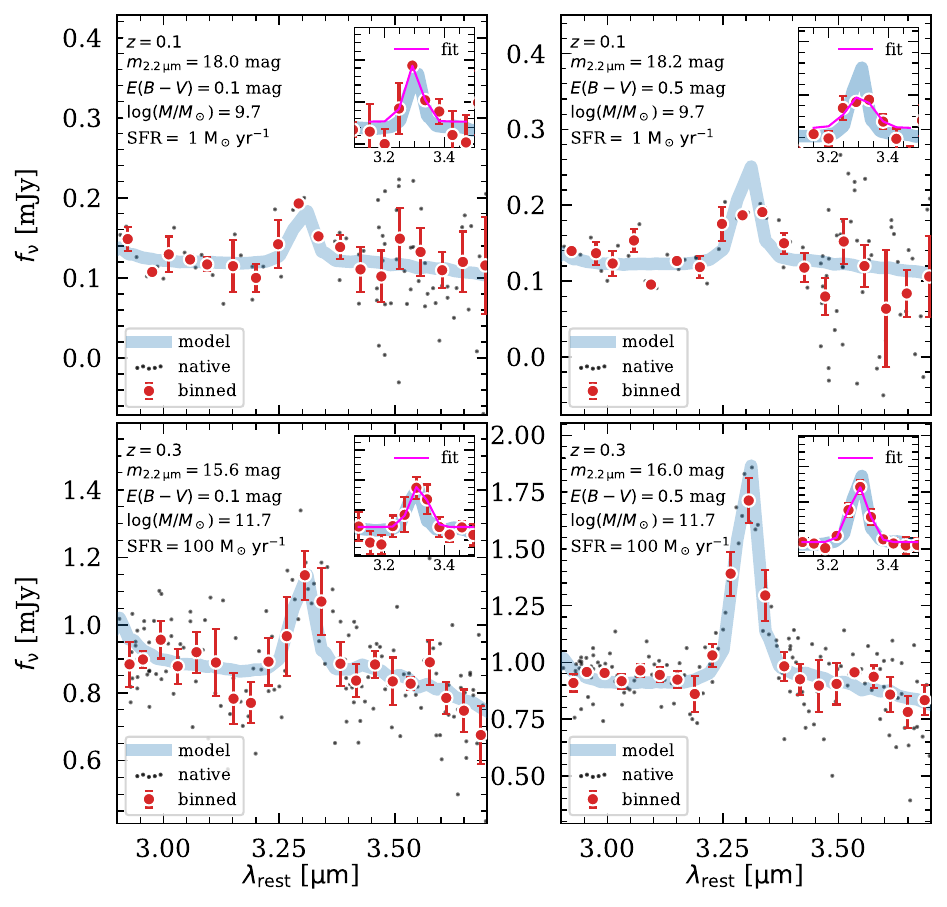}\\
\includegraphics[trim={2.5mm 0 0 0},clip,angle=0,width=1.0\columnwidth]{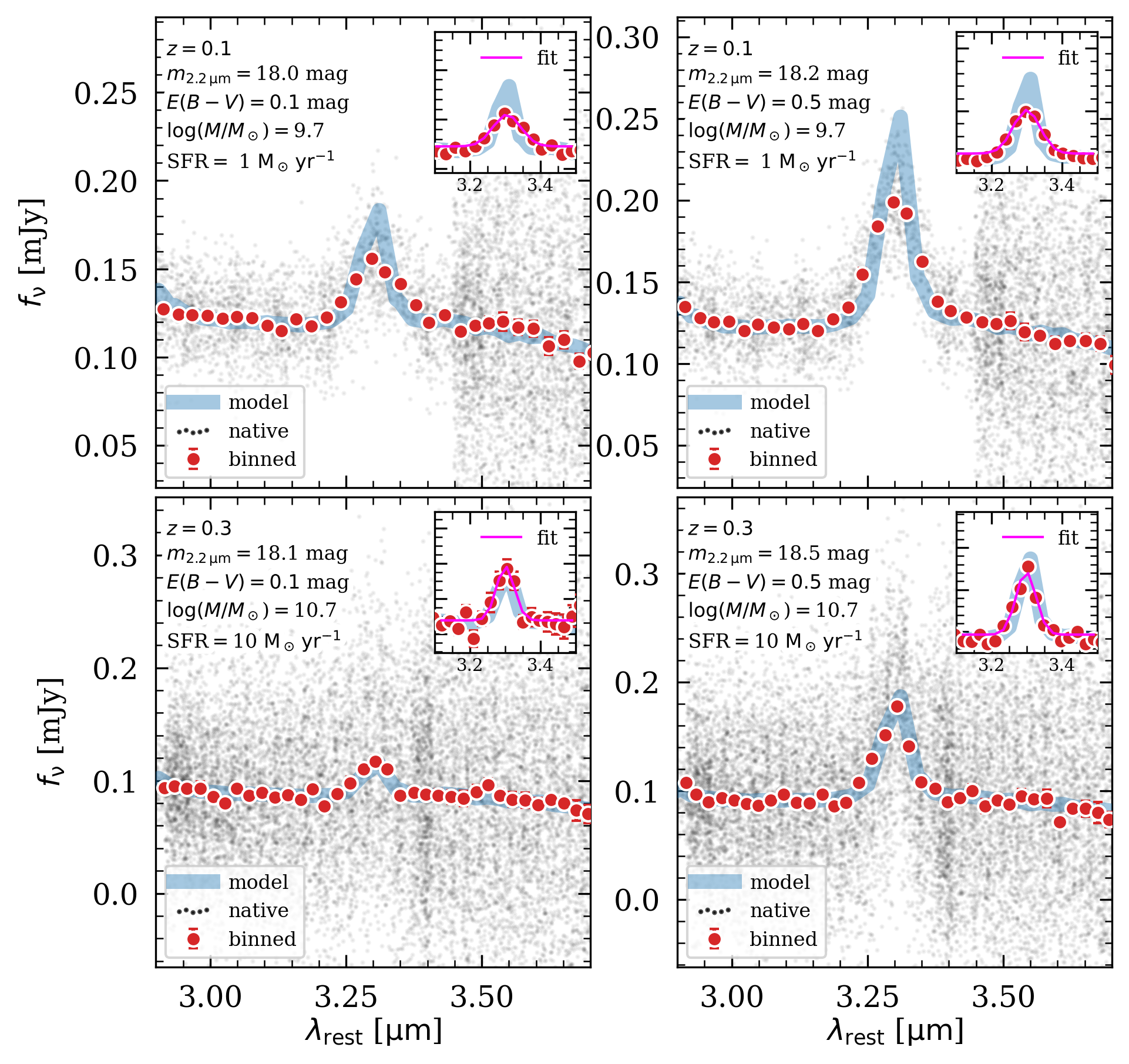}\\\vspace{-2mm}
\caption{Examples of simulated spectra for sources in the wide-field area ({\it top four panels}) and the deep-field area ({\it bottom four panels}). The two areas differ in the number of scans per pixel (see Section~\ref{sec:spherexsim}) and thus number of observations (gray points). The sources are chosen to have a detected PAH $3.3\,{\rm \mu m}$ band emission at $z=0.1$ and $z=0.3$ bracketing \ebmv~values of $0.1$ and $0.5\,{\rm mag}$.
The input model is shown as a thick blue line, and intrinsic measurements are shown as gray points. We also show the measurements binned at $0.1\,{\rm \mu m}$ (wide field) and $600\,{\rm \AA}$ (deep field) in red (errors computed from bootstrapping)}. The insets show the wavelength around the PAH $3.3\,{\rm \mu m}$ band, as well as a fit (magenta line) to the binned data. We note that the significant increase in scatter around $3.45\,{\rm \mu m}$ rest frame ($z=0.1$) is due to a lower sensitivity in detectors $5$ and $6$.
\label{fig:simulationexamples}
\end{figure}
%%%%%%%%%%%%%%%%%%%%%%%%%%%%%%%

\section{Simulations of Nearby Galaxies}\label{sec:sims}

\subsection{Parameter Space of Simulated Spectra}\label{sec:templates}

We simulate spectra to probe evenly a realistic parameter space spanned by redshift, SFR, \ebmv~, and the observed $2.2\,{\rm \mu m}$ ($K-$band) magnitude. We use the distribution of sources from the HSC AKARI NEP photometry catalog \citep{kim21} as motivation for the grid. Specifically, we simulate sources at redshifts $z=0.1-0.4$ in steps of $\Delta z = 0.1$, with dust attenuation values of \ebmv~$=0.05$, $0.1$, $0.3$, and $0.5\,{\rm mag}$, and SFR of $0.01$, $0.1$, $1$, $10$, and $100\,{\rm M_\odot\,yr^{-1}}$. This results in stellar masses between $4\times10^{7}\,{\rm M_\odot}$ and $5\times10^{11}\,{\rm M_\odot}$. The corresponding $K$--band magnitudes depend on redshift and bracket $13.0-23.0\,{\rm mag}$ ($z=0.1$) and $15.5-25.5\,{\rm mag}$ ($z=0.4$).

%%%%%%%%%%%%%%%%%%%%%%%%%%%%%%%

\subsection{Simulation of Realistic Spectra with the \spherex~\texttt{QuickCatalog} Simulator}\label{sec:spherexsim}

We now obtain realistic observed calibrated model \spherex~spectra using the \texttt{QuickCatalog} module of the \spherex~{\it Sky Simulator} \citep{crill20}. The \texttt{QuickCatalog} module includes the realistic noise properties of the \spherex~detectors, as well as galactic and zodiacal background emission. It also uses a realistic point spread function and simulates cosmic-ray hits by masking random pixels on the detector. However, in the following we do not simulate spacecraft jitter (due to pointing stability and slewing motions), as we expect this to be less than $1$ pixel in size and therefore to not have a significant effect on our results. Uncertainties on the measurements are derived from a Monte Carlo sampling over many observations. Furthermore, it is worth noting that the simulator assumes that all sources are point sources and no blending of sources is simulated.
The point-source assumption is true for most galaxies even at lower redshifts, due to the large pixel size ($6.15\arcsec/{\rm px}$) and PSF ($>3\arcsec$). The simulator therefore assumes that the apertures used for the flux measurement are chosen to be large enough to enclose the total flux of the source.
%(which is reasonable for estimated PSF sizes of $>3\arcsec$) and no blending of sources is simulated.
The \spherex~nominal $2$ yr all-sky mission will follow an observing plan in which the spacecraft is scanning the sky in the north-south direction. This stepping is necessary because the data are taken in exposures of linear variable filters (LVFs); thus, each pixel on the image in scanning direction has a different wavelength\footnote{See \url{http://spherex.caltech.edu/} for visualizations of the observations.}. Each location on the sky (``wide-field'' area) is covered by approximately four exposures during the mission. However, because the scans overlap at both the NEP and SEP, those fields are each covered by $\sim 50-100\times$ more exposures (``deep field'' area).
This has implications for the signal-to-noise (S/N) of the observations -- depending on the coordinates of the source, the depth and wavelength resolution may change.
In the following, we run the {\it Sky Simulator} for two fixed locations to bracket the extremes: {\it (i)} the COSMOS field \citep{scoville07} at $\alpha = 150^{\circ}$ and $\delta = +2^\circ$ (part of the wide-field area, receiving on average four visits) and {\it (ii)} the median location of the SPICY sources at the NEP at $\alpha = 269.31^{\circ}$ and $\delta = +66.47^\circ$ (part of the inner $r=1.5\,{\rm deg}$ part of the deep-field area, receiving on average $400$ visits).

%%% FIGURE 3 %%%%
\begin{figure}[t!]
%\vspace{5cm}
\centering
\includegraphics[angle=0,width=0.45\textwidth]{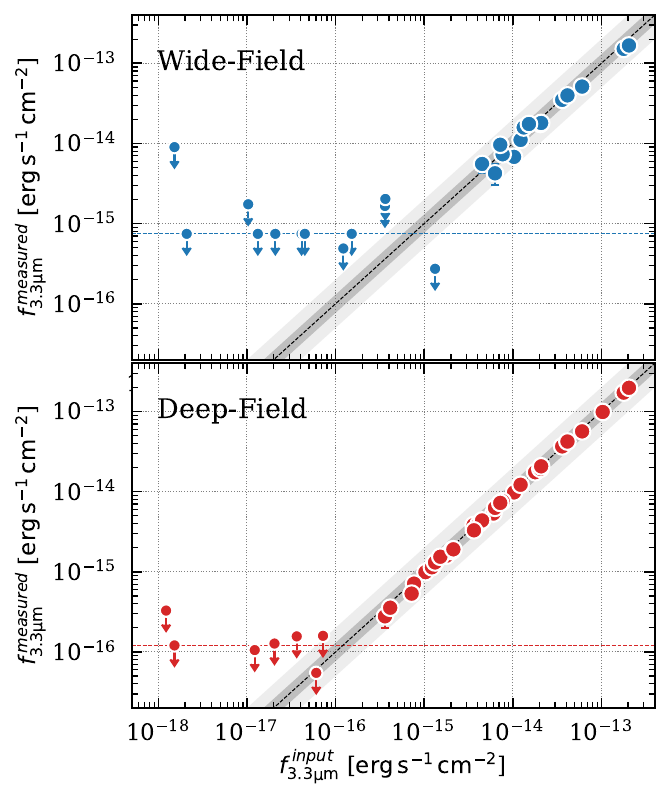}\\\vspace{-3mm}
\caption{Comparison of input and measured PAH $3.3\,{\rm \mu m}$ total flux for the wide-field ({\it top panel}) and deep-field ({\it bottom panel}) survey configuration. The gray areas show an error of $30\%$ (dark gray) and a factor of two (light gray) around the 1-to-1 line (dashed). The dashed horizontal lines show the median $1\sigma$ detection limits.
\label{fig:fluxcomparison}}
\end{figure}
%%%%%%%%%%%%%%%%%%%%%%%%%%%%%%%

Figure~\ref{fig:simulationexamples} shows examples of the simulated spectra at the end of the 2 yr mission of \spherex. We show four sources with different redshifts, dust attenuation, and SFRs (corresponding to $K$--band magnitude and stellar mass) imaged in the wide-field (top panel) and the deep-field (bottom panel) area. The two simulations differ by the number of observations. The deep-field area contains approximately $100\times$ more observations. This has two effects: the S/N can be increased by stacking more individual observations per bin, and the spectral resolution can be increased by applying narrower sampling bins.
In our example, we chose a bin width of $0.1\,{\rm \mu m}$ for the wide-field observations and $600\,{\rm \AA}$ for the deep-field observations.
This is a factor of $2.5-5\times$ smaller than the expected observed PAH $3.3\,{\rm \mu m}$ FWHM \citep[e.g.,][]{lai20}.
Note that the individual observations (gray points in Figure~\ref{fig:simulationexamples}) are available in the \spherex~{\it primary} catalogs, and the user can apply their own binning scheme. 
The inset in each subpanel shows a Gaussian fit to the binned measurements (red) at the location of the PAH $3.3\,{\rm \mu m}$ band. The PAH emission can be recovered approximately $2.5\,{\rm mag}$ deeper in the deep-field area compared to the wide-field area in the example of a $z=0.3$ source.
In Section~\ref{sec:results}, we discuss more quantitatively the parameter space in which the PAH $3.3\,{\rm \mu m}$ emission can be detected in the two different observation areas.

%\subsection{Generation of Spectral Templates}

%\begin{itemize}
%    \item detail the generation of the spectral templates
%    \item this includes: normalization of the template spectra to the photometry and adding PAH features to it ($3.3\,{\rm \mu m}$ and $3.4\,{\rm \mu m}$, etc)
%\end{itemize}

%\subsection{Create {\it SPHEREx} Simulated Spectra}

%\begin{itemize}
%    \item how the simulator works (from a user perspective): \texttt{QuickCatalog}.
%    \item the back-end with {\it Tractor} extension.
%    \item show some example spectra
%\end{itemize}

%\subsection{Analysis of Simulated Spectra}\label{sec:analysis}

%\begin{itemize}
%    \item Maybe we should have this subsection to outline the measurements we do on the simulated spectra
%    \item this could include the measurement of the PAH line with some fitting program (or simply summing up the flux).
%\end{itemize}

%%% FIGURE 4 %%%%
\begin{figure*}[t!]
%\vspace{5cm}
\centering
\includegraphics[angle=0,width=1\textwidth]{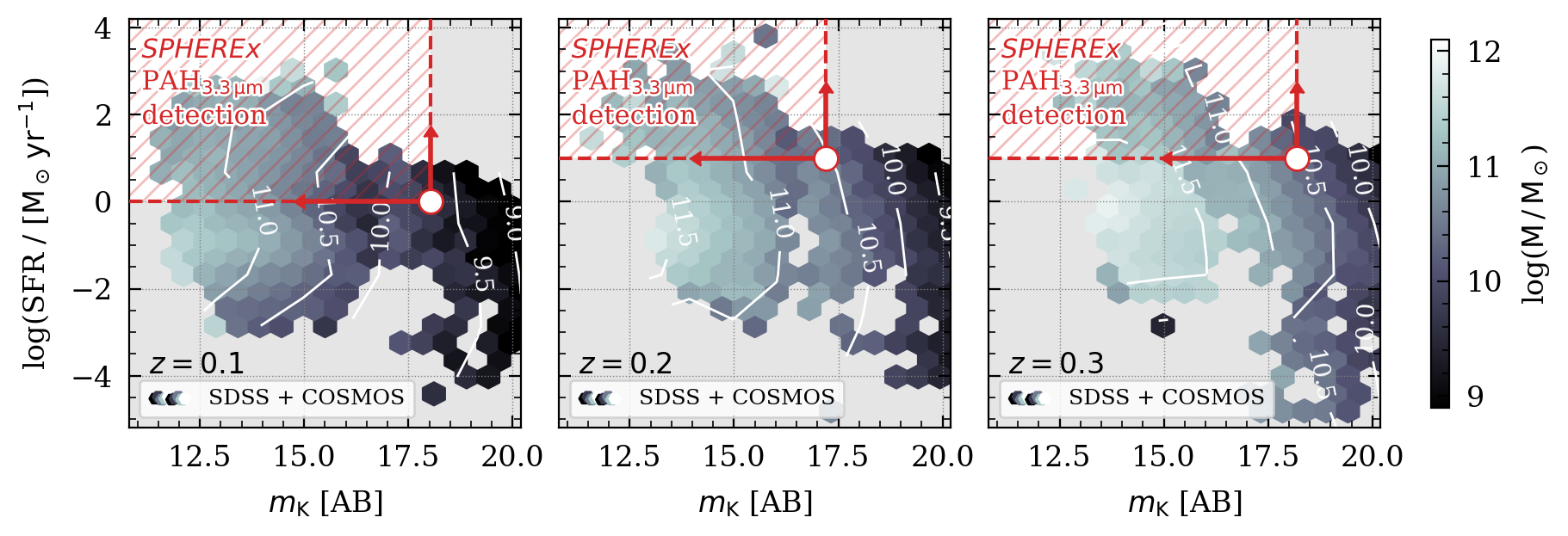}\vspace{-5mm}
\caption{Summary of SFR vs. observed $K$--band  magnitude (proxy for stellar mass; contours) parameter space for three different redshift bins in which \spherex~is able to measure the PAH $3.3\,{\rm \mu m}$ emission band (indicated by the light-red hatched region). The hexagonal background shows galaxies in the COSMOS field and from SDSS in the given redshift bins color-coded by average stellar mass.
\label{fig:paramspace}}
\end{figure*}
%%%%%%%%%%%%%%%%%%%%%%%%%%%%%%%

\section{Results}\label{sec:results}

As described in Section~\ref{sec:templates}, we have created a grid in redshift, SFR, $K$--band magnitude, and stellar mass to access the feasibility of the detection and measurement of the PAH $3.3\,{\rm \mu m}$ emission band.
For each simulated spectrum, we perform a single Gaussian fit to the continuum-subtracted emission line (see Figure~\ref{fig:simulationexamples}) and compare it to the known line flux measured on the input simulated spectrum \footnote{ We note that the PAH emission band is commonly described with a broader Lorentzian profile. However, at the S/N and spectral resolution of \spherex, we find that a Gaussian is a sufficient approximation.}. During fitting, we fix the central wavelength of the Gaussian to $3.3\,{\rm \mu m}$ and let the $\sigma$ and amplitude vary. In this simple calculation, and since we are only interested in the total emission flux, we do not incorporate the \spherex~filter profiles. Furthermore, we find that a Gaussian shape is a good-enough approximation to the $3.3\,{\rm \mu m}$ PAH emission band, as more realistic spectral profile templates would not lead to a significant improvement of the fits.

Figure~\ref{fig:fluxcomparison} compares the input to the measured PAH $3.3\,{\rm \mu m}$ emission band flux (using the method detailed above) for each simulated galaxy in the parameter space described in Section~\ref{sec:templates}.  
Generally, the total emission-line band flux can be recovered to better than $30\%$ (indicated by the dark-gray band).
However, as expected, there is a significant difference in the recovered limit in PAH line flux for galaxies in the wide-field and deep-field areas. In the wide-field survey area (approximately four visits), we find that emission-line fluxes larger than $2\times 10^{-15}\,{\rm erg\,s^{-1}\,cm^{-2}}$ can be recovered to that precision. For the deep-field part of the survey ($\sim100\times$ more exposures), on the other hand, fluxes about $10\times$ fainter ($3\times 10^{-16}\,{\rm erg\,s^{-1}\,cm^{-2}}$) can still be measured robustly.
Note that these line flux sensitivities obtained for the wide- and deep-field observations are in line with what was found in \citeauthor{feder24} (\citeyear{feder24}; e.g., compare with their figure 20).

We note that the above results are for our assumption of a $q_{\rm PAH}$ of $1.12\%$. However, as shown in Figure~\ref{fig:cigalecalibration}, there may be a significant scatter around this value. Specifically, $q_{\rm PAH}$ values in the range of $0.6-1.5\%$ can reproduce the observations from \citet{lai20} at high SFR to which \spherex~will be sensitive to. We note that the measurement accuracy and sensitivity {\it per se} do not depend on the adopted $q_{\rm PAH}$. However, the physical relation between PAH $3.3\,{\rm \mu m}$ flux and physical quantities does. If we assume the observed range in $q_{\rm PAH}$ and further assume the same galaxy population, it would result in a range of line detection thresholds for the wide-field survey area ($2-4\times 10^{-15}\,{\rm erg\,s^{-1}\,cm^{-2}}$) and similarly for the deep-field area ($2-6\times 10^{-16}\,{\rm erg\,s^{-1}\,cm^{-2}}$).

% 0.53 - 1.33
% 1.86 lower - 1.33 higher qPAH

Figure~\ref{fig:paramspace} summarizes the parameter space in SFR vs. observed $K$--band magnitude (or equivalently stellar mass indicated by white contours) for which robust measurements (within $30\%$ error) of the total PAH $3.3\,{\rm \mu m}$ emission band flux are possible with \spherex.
The different panels show the parameter space for the redshift cases of $z=0.1$, $0.2$, and $0.3$. For all cases, we assume an $\ebmv = 0.1\,{\rm mag}$. This will result in conservative limits, as at these high stellar masses a high dust attenuation is expected \citep[e.g.,][]{calura17}. We note that galaxies with a higher $\ebmv$ value can be detected to lower SFR and stellar masses (or $K$--band magnitudes) by a factor $\sim2-4$.
We find that for all redshifts the limiting observed $K$--band magnitude is $16-18\,{\rm mag}$ for which we can derive reliable PAH $3.3\,{\rm \mu m}$ emission band fluxes. This corresponds to stellar masses of $\logm = 9.5$ for $z=0.1$, $\logm = 10.5$ for $z=0.2-0.3$, and $\logm = 11$ for $z=0.4$. For the star formation, we expect limits at $1\,{\rm M_\odot\,yr^{-1}}$ ($z=0.1$), $10\,{\rm M_\odot\,yr^{-1}}$ ($z=0.2-0.3$), and $100\,{\rm M_\odot\,yr^{-1}}$ ($z=0.4$).
Following up on the discussion above on the variations in the assumed $q_{\rm PAH}$ value, we note that both the stellar mass and SFR limits are {\it increased} by a factor of $1.8$ for a $q_{\rm PAH}$ value of $0.6\%$ and {\it decreased} by a factor of $1.3$ for a $q_{\rm PAH}$ value of $1.5\%$.

This means that \spherex~will be able to survey the PAH emission in relatively massive, star forming galaxies out to $z\sim0.4$. In Figure~\ref{fig:paramspace} we also show the coverage of the parameter space by real galaxies drawn from SDSS \citep{abdurrouf22_sdssdr17} and the COSMOS survey. For the former, we use the stellar masses and SFR derived at spectroscopic redshifts \citep{kauffmann03,brinchmann04,salim07}, as well as the Two Micron All Sky Survey $K-$band photometry \citep{skrutskie06}. For the latter, we use the physical parameters derived in the COSMOS2020 catalog \citep{weaver22}, including the $K_{\rm s}$ photometry from UltraVISTA \citep{mccracken12}. 
Note that at $z=0.4$, only COSMOS data are available as there are no SDSS measurements of stellar mass and SFR that are measured in a consistent way to the lower redshift bins. We therefore do not show this redshift bin in Figure~\ref{fig:paramspace}.
We then use these derived limits in stellar mass, SFR, and observed $K-$band as shown in Figure~\ref{fig:paramspace} to determine the approximate number of galaxies over the survey area of \spherex. We do this by extrapolating the number counts from the sky coverage of SDSS ($\sim 14,000\,{\rm deg^2}$) to the sky coverage of \spherex~($\sim 28,000\,{\rm deg^2}$ after subtracting the galactic and ecliptic planes). We find the expected number of sources within this parameter space to be at least $100\,000$ for $z=0.1$, $10\,000$ for $z=0.2$, and $4\,000$ for $z=0.3$ across the whole sky. The extrapolation of the stellar mass functions derived on the COSMOS field \citep{davidzon17} suggests that several hundred sources will be detected at $z=0.4$ in this given parameter space covered by \spherex.

%\begin{itemize}
%    \item simulate spectra for different broad-band magnitudes and SFR (which affect the PAH emission). 
%    \item analyze how well we can measure PAH features. (we need some kind of analytical way to define what ``good'' means in that sense. Maybe a $\chi^2$ of the fit or a S/N would make sense.
%    \item make some predictions:
%    {\it (i)} what is the SFR/mass/magnitude range in which we can successfully measure PAH features for the wide and deep survey?
%    {\it (ii)} is the resolution (in wide and deep) enough to measure certain spectral features (e.g. Br$\alpha$ or differentiate between the PAH $3.3\,{\rm \mu m}$ and aliphatic $3.4\,{\rm \mu m}$ feature?
%    {\it (iii)} an estimate of the number density (or total number of source) where the PAH can be measured?
%    {\it (iv)} some implications on the mid-IR continuum measurements?
%    {\it (v)} what about resolved measurements? We could actually simulate a \spherex~image of a (large) source (e.g., nearby galaxy from the {\it GOALS} sample) and show how well we can resolve the PAH feature in the galaxy's disk.
%\end{itemize}

\section{Conclusions}\label{sec:conclusions}

During its nominal $2$ yr mission, \spherex~will construct an all-sky spectral map from $0.75 - 5\,{\rm \mu m}$ observations of stars, galaxies, and diffuse emission. Due to its survey strategy, \spherex~will observe the deep fields (SEP and NEP) at approximately $100\times$ higher cadence, resulting in order-of-magnitude-deeper observations.
In this work, we investigated the potential of \spherex~to detect the PAH $3.3\,{\rm \mu m}$ emission-line band in galaxies out to $z=0.4$ after the $2$ yr mission.

To this end, we used the publicly available SED modeling code CIGALE to construct mid-IR galaxy spectra that include the PAH $3.3\,{\rm \mu m}$ emission feature, to assess the potential for \spherex~to detect the PAH feature out to $z=0.4$ using data from its nominal $2$ yr mission. These models were then calibrated with real observations of AKARI spectra and the observed relation between SFR and $3.3\,{\rm \mu m}$ PAH emission over seven orders of magnitudes. We constructed templates to evenly cover the parameter space in redshift, SFR, stellar mass, $K$--band magnitude, and $\ebmv$. These templates were run through the \spherex~simulator to obtain realistic observed spectra.

From these simulations, we find that \spherex~will be able to measure the PAH $3.3\,{\rm \mu m}$ emission band total fluxes to better than $30\%$ down to $3\times10^{-15}\,{\rm erg\,s^{-1}\,cm^{-2}}$ in the wide-field part of the survey and approximately one order of magnitude deeper ($3\times10^{-16}\,{\rm erg\,s^{-1}\,cm^{-2}}$) in the deep-field parts. This translates into the parameter space summarized in Figure~\ref{fig:paramspace}. In terms of SFR vs. stellar mass, we expect to measure the PAH $3.3\,{\rm \mu m}$ emission in several hundred thousand galaxies across the sky at $z=0.1$ down to $\logm = 9.5$ and an SFR of $1\,{\rm M_\odot\,yr^{-1}}$. For $z=0.2-0.3$, we expect a few thousand down to $\logm = 10.5$ and an SFR of $10\,{\rm M_\odot\,yr^{-1}}$. At higher redshifts, the number densities drop owing to an increased SFR and stellar mass sensitivity limit, but still a few hundred  sources are expected.

\spherex~will be transformative not only in Milky Way science and cosmology but also in the study of dust via the PAH $3.3\,{\rm \mu m}$ spectroscopic feature in extragalactic systems. It will increase the number of galaxies with such measurements from earlier works with Spitzer, AKARI, and JWST by a factor of $10\,000$ and therefore allow a population study of dust grain properties in different environments and for different galaxy types, such as star-forming, AGN, or even quiescent galaxies, by stacking.

%% Also note that the akcnowlodgment environment does not support long amounts of text. If you have a lot of people and institutions to acknowledge, do not use this command. Instead, create a new \section{Acknowledgments}.
\begin{acknowledgments}
{\it Acknowledgements:} We thank the anonymous referee for their valuable input, which improved this manuscript substantially.
We thank the \spherex~science team for valuable discussions and support in writing this paper.
E.Z. acknowledges the support of the {\it Soli Deo Gloria} SURF Fellow Donors.
Y.K. was supported by the National Research Foundation of Korea (NRF) grant funded by the Korean government (MSIT; No. 2021R1C1C2091550).
H.I. acknowledges support from JSPS KAKENHI grant No. JP21H01129.
Part of this work was done at Jet Propulsion Laboratory, California Institute of Technology, under a contract with the National Aeronautics and Space Administration (contract No. 80NM0018D0004).
Based on observations collected at the European Southern Observatory under ESO programmes 179.A-2005, 198.A-2003, 1104.A-0643, 110.25A2, and 284.A-5026; and on data obtained from the ESO Science Archive Facility \citep{data_ultravista}; and on data products produced by CANDIDE and the Cambridge Astronomy Survey Unit on behalf of the UltraVISTA consortium.
\end{acknowledgments}


\vspace{5mm}
\facilities{ {\it AKARI}, {\it Sloan}, {\it ESO:VISTA} }


\software{\texttt{CIGALE} \citep{burgarella05,noll09,boquien19}; \spherex~Sky Simulator \citep{crill20}; \texttt{astropy} \citep{astropy13,astropy18}}



\appendix
\section{Dependence on Different Model Assumptions}\label{app:differentmodel}

In the main part of this paper, we have assumed a constant SFH. However, we have seen that most galaxies that \spherex~will detect are highly star forming. While they can still be on the star-forming main sequence, some of them are surely starbursts and may even contain an AGN. Here we investigate the effect of different model assumptions on the derived PAH $3.3\,{\rm \mu m}$ flux.

Specifically, we assume three more models, including
\begin{itemize}
    \item a constant SFH with exceptionally high dust attenuation of $\rm E(B-V) = 1\,{mag}$;
    \item a starburst model (assuming a $200\,{\rm Myr}$ long period of $10\times$ increased SFR) for a low ($0.1\,{\rm mag}$) and high ($1.0\,{\rm mag}$) $\rm E)B-V)$ case; and
    \item a model with a dust-obscured AGN contributing $80\%$ to the mid-IR emission.
\end{itemize}

\begin{figure*}[!h]
%\vspace{5cm}
\centering
\includegraphics[angle=0,width=0.55\textwidth]{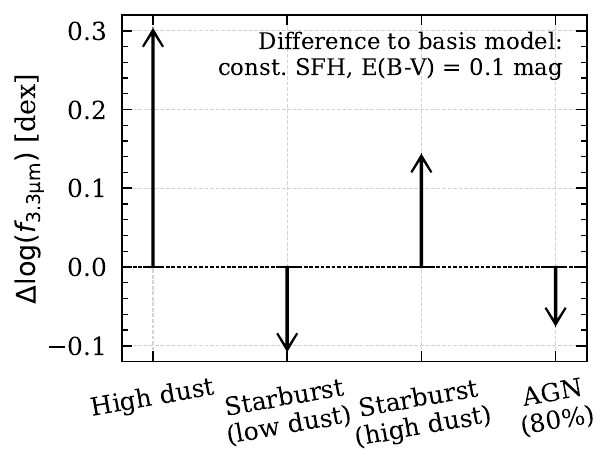}\vspace{-5mm}
\caption{Impact of different model assumptions (SFH, dust, and AGN) on the PAH $3.3\,{\rm \mu m}$ emission flux (resulting change with respect to the basis model in dex).
\label{fig:differentmodel}}
\end{figure*}

Figure~\ref{fig:differentmodel} shows the resulting change in the PAH $3.3\,{\rm \mu m}$ flux for these different model implementations. The differences are calculated with respect to a fiducial basis model, which assumes a constant SFH and a dust attenuation of $\rm E(B-V) = 0.1\,{mag}$.
It can be seen that the change $\rm E(B-V) = 0.1 \rightarrow 1.0\,{mag}$ increases the PAH emission by $0.3\,{\rm dex}$ (a factor of two).
As expected, the inclusion of an AGN decreases the PAH emission, as its output radiation leads to the destruction/dissociation of small dust grains. However, we note that this only happens in the core of the galaxy in close proximity of the AGN, while the disk still shows PAH emission at a nominal level \citep[e.g.,][]{lai23}. Therefore, this decrease may be larger than actually observed.
The case of the starburst is more complicated and interesting. The PAH emission of a starburst is naturally increased owing to the increased SFR (see  Figure~\ref{fig:cigalecalibration}). To remove this dependency and to focus on the impact of the different SFH alone, we compare the starburst model to a constant SFH model of equivalent current star formation. After this, we find a {\it decrease} in PAH emission comparing to the same dust attenuation. This may be because dust has not had time to build up over the short duration of the starburst. Note that a longer starburst approaches the values of a constant SFH. However, starbursts are usually accompanied by dust-obscured star formation. It is therefore likely that the effective impact of a starburst is an increase in the PAH emission (as shown in Figure~\ref{fig:differentmodel} in the case of $\rm E(B-V) = 1\,{\rm mag}$).

Overall, we find that different SFHs and the inclusion of AGN do not significantly impact our results by more than a $0.1-0.15\,{\rm dex}$ ($25-50\%$) change in PAH emission. This would translate into an equally small factor in the sensitivity calculation and hence the limiting SFR and stellar mass.

%%% FIGURE 5 %%%%
%%%%%%%%%%%%%%%%%%%%%%%%%%%%%%%


\bibliography{bibli}
\bibliographystyle{aasjournal}



\end{document}